\documentclass[a4paper]{article}

\usepackage{marginnote}

\usepackage{INTERSPEECH2022}

\title{Attack Agnostic Dataset: Towards Generalization and Stabilization of Audio DeepFake Detection}

\name{Piotr Kawa$^1$, Marcin Plata$^1$, Piotr Syga$^1$}

\address{
  $^1$Wroc{\l}aw University of Science and Technology, Wroc{\l}aw, Poland
}
\email{\{piotr.kawa, marcin.plata, piotr.syga\}@pwr.edu.pl}

\begin{document}

\maketitle

\begin{abstract}
    Audio DeepFakes allow the creation of high--quality, convincing utterances and therefore pose a threat due to its potential applications such as impersonation or fake news. Methods for detecting these manipulations should be characterized by good generalization and stability leading to robustness against attacks conducted with  techniques that are not explicitly included in the training. In this work, we introduce Attack Agnostic Dataset --- a combination of two audio DeepFakes and one anti--spoofing datasets that, thanks to the disjoint use of attacks, can lead to better generalization of detection methods. We present a thorough analysis of current DeepFake detection methods and consider different audio features (front--ends). In addition, we propose a model based on LCNN with LFCC and mel--spectrogram front--end, which not only is characterized by a good generalization and stability results but also shows improvement over LFCC--based mode --- we decrease standard deviation on all folds and EER in two folds by up to 5\%.
\end{abstract}


\section{Introduction}

Recent advancements in the field of neural networks, provide better results in many signal processing tasks compared to using traditional methods. 
One of the effects of such development is a large number of modified images, videos and audio, including the phenomena known as DeepFakes (DF). 
The origin of DF was substituting the original face in the video with the victim's face for a humorous purpose while using the same technique may seriously affect the victim --- from besmearing and fake news to impersonation in biometric authentication. 

Due to the threats associated with DeepFakes, a significant body of work has been dedicated to detecting the manipulated records. Due to the detection task's considerable difficulty and the carrier's popularity, most research focus on visual manipulation. However, recent years have been marked by vast enhancements in the field of speech manipulation and speech generation. 
The Audio DeepFakes involve changing the original voice on a recording to sound like the victim or generating new utterances with the use of someone's voice. Due to this fact, the area is similar to the problem of spoofing, voice splicing and frequency (or spectrogram) analysis.
There has been a vast improvement in the topics utilized by DeepFake creation methods: Text--To--Speech (TTS)~\cite{tacotron-2},
vocoders~\cite{waveglow,wavenet,wavernn,melgan}, Voice Conversion~\cite{stargan_first} and Voice Cloning~\cite{rtvc}. These new solutions, mostly based on deep neural networks, generate high--quality utterances that achieve high results in mean opinion scores evaluated by human listeners. The improvements leading to the risks associated with the less common audio DeepFakes, are not to be ignored as producing fake messages, impersonation~\cite{audio-scam} or surpassing speaker recognition systems may result in severe consequences for the victim.

There exists a variety of audio spoofing detection methods. Some of them has been adapted to the topic of DeepFake detection.
One of the most common approaches is based on Gaussian Mixture Models (GMM)~\cite{gmm}.
Recent solutions have been based on deep neural networks, e.g. convolutional neural networks (CNNs)~\cite{lcnn,selcnn}. 
Another group of methods is composed of raw audio signal analysis ~\cite{rawnet2,rawgat-st}.
Recently, vision DF detection methods have been adapted to the field of audio analysis~\cite{fakeavceleb}.

Neural networks (as well as other machine learning models) applied to the DeepFake detection problem are particularly prone to two common issues --- generalization and stability. Unlike other audio tasks (e.g. speech detection or speaker recognition), the DeepFake detection models are trained to find some hidden artifacts and distortions since the general characteristics of audio samples for bona fide and spoof are similar.
Moreover, in the general case, these artifacts are not known to human experts. Thus from the perspective of a loss function used to train the neural network, we are often limited to the binary categorization (spoof vs. bona fide).

Audio DeepFakes are phenomena similar to another, much older field of tampering --- audio spoofing. It includes methods like replay attacks and voice conversion of voice synthesis.
The main difference between these two fields is their application and target to be deceived. Spoofing is used to fool automatic speech verification methods in determining if a given voice belongs to a certain person. DeepFakes are used to deceive humans --- algorithms aim to convince someone that a given sample is bona fide. In addition --- DF manipulations always generate new samples, whereas spoofing can be composed of existing utterances or be a replayed sample.

The problem of the generalization has been noticed recently for the DeepFake detection task (and anti--spoofing). In ASVspoof Challenge~\cite{WANG2020101114,Delgado2021ASVspoof2A}, the train, test and eval subsets were prepared using different attacks settings and in a way that promotes \textit{on the board} models that generalize well. In~\cite{Bialobrzeski2019}, authors proposed \textit{attack--out cross--validation} procedure for replay attacks and split training and testing subsets with respect to microphone distance and replay--device quality. In~\cite{wavefake}, a limited approach was presented, in which only one DeepFake method was excluded from the training set or only one method was included in the training set (i.e. the two extreme cases).
Contrary to the generalization studied for the audio DeepFake problem, stability is often overlooked and underestimated. In our paper, we focus on both of these issues. Our codebase related to this research can be found on GitHub: https://github.com/piotrkawa/attack-agnostic-dataset.

Our contribution in this paper includes:
\begin{itemize}
    \item proposing a new way of evaluating the generalization of models --- Attack Agnostic Dataset,
    \item testing our solution with respect to the stability, the generalization, and ultimately the performance, 
    \item comparison of multiple front--ends' performance and generalization,
    \item introduction of LCNN model trained on the double front--end --- LFCC and mel--spectrogram, that stands out from other state--of--the--art methods in the terms of generalization, stability, and performance.   
\end{itemize}

\section{Datasets}

Due to the fact that DeepFakes are generated using many kinds of methods, it is crucial for the detection model to generalize well, i.e. the solution should be robust to attacks or distortions occurring in a training set as well as to new unknown manipulations and therefore focus on general artifacts which separate bona fides from fakes. 

Another critical case is the stability of the detection model. Most of the recent DeepFakes are generated using neural networks; hence there is no ''list'' of artifacts that could occur in a sample, and it is impossible to check if the model covers all possible artifacts that indicate a fake. One approach to handle the stability is calculating the performance across different fake generating methods and checking if the results are comparable.

\subsection{Sub--datasets}

To enhance these methods' generalization and stability, we introduce Attack Agnostic Dataset. It is a combination of two DeepFake datasets --- FakeAVCeleb~\cite{fakeavceleb} and WaveFake~\cite{wavefake}, and one dataset originally prepared for spoofing detection --- ASVspoof 2019 LA subset~\cite{asvspoof2019}. 

FakeAVCeleb dataset~\cite{fakeavceleb} addresses the shortage of multimodal databases --- containing both visual and audio DeepFakes. The authors used a few approaches to generate a DeepFake video. On the other hand, audio manipulations were obtained only using one algorithm SV2TTS~\cite{rtvc} which is a transfer learning--based real--time voice cloning tool. The first step of voice cloning was to use the IBM Watson speech--to--text service~\cite{ibm_tts} to read a spoken sentence and then SV2TTS was applied to generate the same sentence with cloned voice. The authors noted that this kind of modification was proposed in their dataset for the first time. We use solely audio tracks of the dataset obtaining 500 bona fide and 11,357 fake utterances.

WaveFake~\cite{wavefake} is one of the most recent audio DeepFake datasets. It is created basing on two languages (English and Japanese) with the use of 7 spoof generating neural network architectures: MelGAN (with MelGAN Large)~\cite{melgan}, ParallelWaveGAN~\cite{parallel-wavegan}, Multi--band MelGAN, (with Full--band MelGAN)~\cite{mb-melgan}, HiFi--GAN~\cite{hifi-gan} and WaveGlow~\cite{waveglow}.
All solutions except WaveGlow are based on Generative Adversarial Networks (GANs)~\cite{gan}. Additionally, some of the architectures are variants of others, e.g. MelGAN Large is a version of MelGAN containing bigger receptive field whereas Full--band MelGAN is a variant of MB--MelGAN --- architectures differ in the way their auxiliary loss function is computed.
Our Attack Agnostic Dataset includes 18,100 bona fide samples from LJSpeech~\cite{ljspeech17} and JSUT~\cite{jtsu} databases along with 101,700 fake samples created by WaveFake authors.

ASVspoof~\cite{asvspoof2019} is one of the most notable challenges in audio spoofing detection. Our dataset includes Logical Access subset from 2019 installment --- 12,483 bona fide and 108,978 fake samples. Attacks were created using text--to-speech (TTS) and voice conversion (VC) algorithms. In summary, there are proposed 19 methods. The dataset of the most recent installment, ASVspoof 2021~\cite{asvspoof2021}, comprises of 3 databases --- logical access (LA), physical access (PA) and a new subset --- speech DeepFake\footnote{We thank Xin Wang from National Institute of Informatics, Tokyo for pointing out the mistake in ASVspoof dataset description.}.
Please note that we did not use this dataset as at the time of the paper being written, DF subset training information, as well as, LA attack types, were not available. The information about the attack types is crucial for our approach.

Since spoofing methods aim at deceiving automatic speaker verification systems (contrary to human perception as in DeepFakes), the methods used to generate spoofs are slightly different. Therefore, artifacts generated by spoofing methods could also differ from DeepFakes. For example, spoofing methods could generate some crackling noises (as in a headphone/handset) to confuse the ASR, while similar noises returned by DeepFakes methods may induce suspicion in a listener. For the sake of the best generalization possible, we consider both types of artifacts, and for that reason, we combine the LA subset of ASVspoof dataset with DeepFakes datasets.

In addition, note that DeepFake detection databases often lack diversity in terms of ethnicity or age. These issues result in a lack of universal datasets that consider a broad spectrum of manipulation methods and allow the generalization of models trained on them. Combining several independent datasets increase the general diversity of the combined set.

\begin{table*}[th]
  \caption{Results of main architectures evaluated on eval set --- all solutions, apart from RawNet2, are based on LFCC front--end. We report the mean and standard deviation of EER (\%) computed for each dataset fold, averaged across 3 randomness seeds.}
  \label{tab:main-architectures}
  \centering
  \begin{tabular}{ c c c c c c c c c c c }
    \toprule
    
    \multicolumn{1}{c}{\textbf{Fold}} & 
    \multicolumn{2}{c}{\textbf{LCNN}} &
    \multicolumn{2}{c}{\textbf{XceptionNet}} & 
    \multicolumn{2}{c}{\textbf{MesoInception--4}} &
    \multicolumn{2}{c}{\textbf{RawNet2}} &
    \multicolumn{2}{c}{\textbf{GMM}} \\
    & EER & STD & EER & STD & EER & STD & EER & STD & EER & STD \\
    
    1 & 9.526 & 0.728 & 16.206 & 1.476 & 38.581 & 6.660 & 19.443 & 1.358 & 25.137 & 0.723 \\ 
    2 & 9.523 & 0.720 & 12.766 & 1.171 & 36.381 & 5.936 & 23.793 & 2.046 & 34.382 & 1.440 \\ 
    3 & 2.370 & 0.493 & 13.056 & 3.224 & 17.646 & 5.622 & 14.144 & 0.891 & 25.176 & 0.667 \\
    \bottomrule
  \end{tabular}
\end{table*}

\begin{table*}[th]
\begin{center}
  \caption{Influence of different front--ends (LFCC, MFCC, spectrogram) applied to LCNN architectures when tested on eval set. In addition, we examine the concatenation of the aforementioned features. Values were obtained similarly to Table~\ref{tab:main-architectures}.}
  \label{tab:frontend-comparison}
  \centering
  \begin{tabular}{ c c c c c c c c c c c c c}
    \toprule
    \multicolumn{1}{c}{\textbf{Fold}} & 
    \multicolumn{2}{c}{\textbf{LFCC}} & 
    \multicolumn{2}{c}{\textbf{MFCC}} & 
    \multicolumn{2}{c}{\textbf{MFCC+Spec}} &
    \multicolumn{2}{c}{\textbf{LFCC+Spec}} &
    \multicolumn{2}{c}{\textbf{MFCC+LFCC}} &
    \multicolumn{2}{c}{\textbf{Spec}} \\
    & EER & STD & EER & STD & EER & STD & EER & STD & EER & STD & EER & STD \\
    \midrule
    1 & 9.526 & 0.728 & 15.296 & 0.138 & 15.023 & 0.484 & 9.137 & 0.321 & 13.437 & 1.277 & 38.698 & 1.609 \\
    2 & 9.523 & 0.720 & 8.732 & 0.419 & 7.552 & 0.596 & 9.158 & 0.639 & 8.262 & 0.516 &	30.441 & 1.379 \\
    3 & 2.370 & 0.493 & 5.066 & 0.616 &	4.722 & 0.238 & 3.125 & 0.335 & 3.184 & 0.715 &	30.092 & 0.489\\
    \bottomrule
  \end{tabular}
\end{center}
\end{table*}

\subsection{Attack Agnostic Dataset}

We propose Attack Agnostic Dataset~combining 3 audio DeepFakes and spoofing datasets --- ASVspoof (LA), WaveFake and FakeAVCeleb (audio subset) that contain in total 31,083 real and 222,035 fake samples covering 27 various methods. 

Our aim is to provide a tool that analyzes the model's generalization and stability. Hence, we propose three different splits of the dataset (folds), where DeepFake methods and spoofing attacks are distributed across train, test and eval subsets in a different manner. Modifications used to generate the distortions are disjointly divided across these subsets. This approach allows to point out the differences in the results for individual folds. Significant variations in the final results could indicate that the trained model does not generalize well, and the choice of this model for the final system increases the chance of failure for any new attack. Furthermore, an analysis of the training process (e.g. accuracy after every epoch) could indicate the stability characteristics. Such an approach allows verifying if the accuracy results on the test subsets for particular modifications grow in a monotonic manner, as in the generalization problem, significant variations of the results could be alarming and could indicate that the model is learning new artifacts in every epoch, instead of accumulating knowledge about them. Answering these questions allows us to choose a well--generalized and stable detection model and reduce a chance of failure.

For every fold we choose about $70\%$ of attacks for train subsets (5 for WaveFakes and 12 for ASVspoof, $70\%$ samples for FakeAVCeleb). Remaining attacks are evenly divided between test and eval subsets. For every fold we select different combination of attacks. Real samples are divided with the same proportion ($70/15/15$) across three subsets. Our aim is to test many scenarios, thus some folds contain all considered variants of attacks in a train subset (e.g. MB--GAN is in train, and its Full--band variant is in validation), while other folds are constructed to limit information of similar artifacts (e.g only one of five voice cloning methods from ASVspoof dataset is in train).

To avoid the influence of differently prepared data across three sub--datasets, we process samples following WaveFake procedure~\cite{wavefake}. First, all audio files are resampled to 16 kHz mono. Next, we trimm all silences longer than 0.2s. Finally, depending on the audio duration, samples are either trimmed or padded to 4s. Padding is based on repeating a sample and trimming concatenated audio file to a predefined length. We also apply an oversampling method to balance datasets across targets (spoof vs. bona fide) during training.

\begin{table*}[htb]
\begin{center}
  \caption{Comparison of EER (\%) on eval sub--datasets by LCNN with MFCC and LFCC front--ends (results of 1 seed).}
  \label{tab:ablation-study}
  \centering
  \begin{tabular}{ c c c c c c c c c c }
    \toprule
    \multicolumn{1}{c}{\textbf{Architecture}} & 
    \multicolumn{3}{c}{\textbf{ASVspoof}} &
    \multicolumn{3}{c}{\textbf{WaveFake}} & \multicolumn{3}{c}{\textbf{FakeAVCeleb}} \\

    Fold no. & 1 & 2 & 3 & 1 & 2 & 3 & 1 & 2 & 3\\
    \midrule
    LCNN+LFCC & 18.773 & 16.542 & 3.251 & 2.106 & 0.298 & 0.377 & 5.954 & 2.933 & 6.061 \\

    LCNN+MFCC & 13.636 & 16.426 & 6.729 & 24.403 & 0.400 & 0.876 & 7.465 & 6.846 & 7.323 \\
    \bottomrule
  \end{tabular}
\end{center}
\end{table*}

\section{Benchmark}

The following section describes our evaluation of DeepFake detection models' generalization. Our implementation was based on codebase provided by WaveFake authors~\cite{wavefake}. During tests, we used the following solutions: LCNN~\cite{lcnn}, MesoInception--4~\cite{mesonet}, XceptionNet~\cite{xceptionnet}, RawNet2~\cite{rawnet2} and GMM~\cite{gmm}. 
RawNet2 was the only model trained on a raw signal. All other solutions used various front--ends as an input. Our evaluation concerned a combination of the following features: linear-frequency cepstral coefficients (LFCC), mel--frequency cepstral coefficients (MFCC) and spectrogram--based features.  Spectrogram features were created using STFT followed by mel--scaling --- we operate on absolute value and angle of the obtained representation.
We used both mel-- and linear--based features to provide models with features that are audible and also outside of human hearing scope.
For each neural network, following FakeAVCeleb authors, we created front--ends using 25ms (400 samples) Hann window, window shift of 10ms (160 samples) and 512 FFT points. All front--ends were based on 80 coefficients. As a result, each formed a 2D array of size $80\times N$, where $N$ is the number of frames.  

We conducted 3 training procedures for every neural network --- each on a different fold of Attack Agnostic Dataset. Models were trained for 5 epochs with validation on the test set. Each training ended with calculating metrics on the eval set using the model which scored the best results on the test set. GMM's front--ends and training procedure followed~\cite{wavefake}.

Every model was trained using gradient descent based on Adam optimizer~\cite{kingma2014adam} and binary cross--entropy loss function with a batch size of 128. 
Wherever possible, we used previously reported learning rates: $1\cdot 10^{-5}$ for MesoInception--4 and XceptionNet (following FakeAVCeleb), $1\cdot 10^{-4}$ 
for RawNet2 (following ~\cite{wavefake,rawnet2}) and $1\cdot 10^{-4}$ for LCNN (following~\cite{asvspoof2021}, note that we use the RNN variant) and $1\cdot 10^{-3}$ for GMM (following~\cite{wavefake}). 
To ensure that the results are reproducible, we ran each procedure using 3 fixed randomness seeds. 

The comparison covers Equal Error Rate (EER) --- metric known from tasks of DeepFake detection or anti--spoofing, also presented in works like ASVspoof or WaveFake. In addition, we report training and validation accuracies.

\section{Results}

Table~\ref{tab:main-architectures} contains scores obtained by the aforementioned architectures.
LCNN achieves the best generalization, scoring the lowest EER across all dataset folds --- respectively 9.525\%, 9.523\% and 2.370\%. It is worth mentioning that for each architecture, there exists a disproportion between results scored across folds --- all models, apart from XceptionNet, show the lowest EER on the third fold. Please note that the results of LCNN are characterized by the highest consistency scoring the lowest standard deviation, which indicates its best stability.
XceptionNet provides 2nd best mean EER (16.206\%, 12.766\%, 13.056\%), suggesting that some models typically used in visual DF detection may be successfully applied in the audio domain. However, these results are less stable than the ones scored by RawNet2. Despite higher EER of 19.443\%, 23.793\% and 14.144\%, RawNet2 has lower std of 1.358\%, 2.046\%, 0.891\% in contrary to XceptionNet's 1.476\%, 1.171\%, 3.224\%.

Apart from evaluating different architectures, we have conducted in--depth research regarding the influence of commonly used front--ends on the generalization and stability of Audio DeepFake detection. For this evaluation, we have chosen the LCNN --- model, which provided the best results in the first part of our benchmark. The results are shown in Table~\ref{tab:frontend-comparison}. They are less unambiguous than the ones presented in Table~\ref{tab:main-architectures} --- there is not only a difference in results for each fold but also no universally superior front--end, i.e. no solution presents the lowest EER across all of the concerned folds.
However, treating LFCC as a baseline, each of the other front--ends (apart from spectrogram) provides improvements on at least one fold. Note that the results of LFCC and LFCC+Spectrogram are the most similar, and the later improves EER in case of 2 out of 3 folds. We point out that LFCC+Spectrogram is also characterized by a better stability scoring std of 0.321\%, 0.639\% and 0.335\%.
High dissimilarities across all folds show that MFCC is the most unstable front--end and that its results highly depend on the data. This feature provides worse results than LFCC; moreover, concatenation of MFCC and LFCC, despite improvement in relation to MFCC, still provides results worse than LFCC.
Spectrogram features present the worst results across all front--ends; however attaching it to other features enhances the results (apart from the fold 3 of LFCC+Spectrogram).

\begin{figure}[t]
  \centering
  \includegraphics[width=\linewidth]{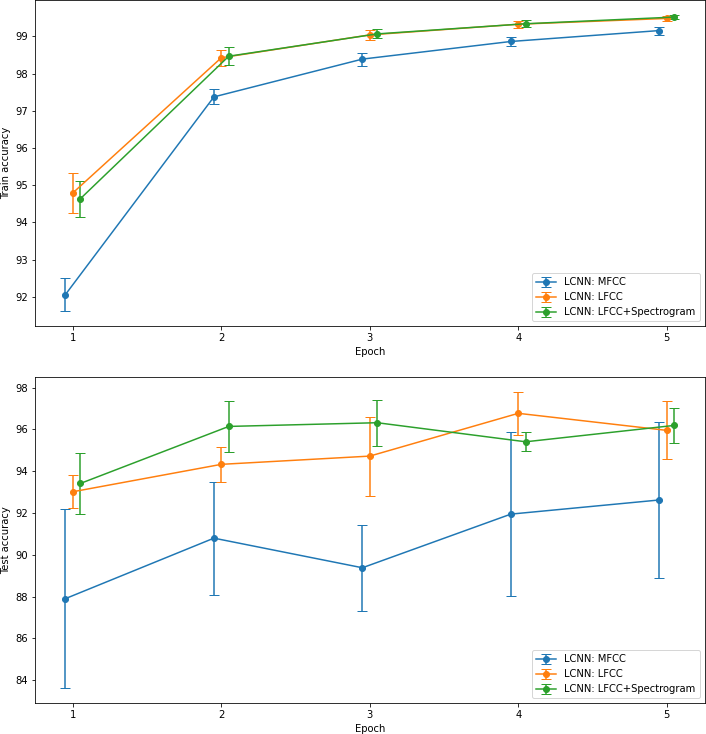}
  \caption{Accuracy results along with their standard deviations for LCNN with three selected front--ends calculated on train (upper) and test (lower) subsets.}
  \label{fig:stability_train_test}
\end{figure}

\section{Ablation study and Discussion}

We observe, according to the results presented in Table~\ref{tab:ablation-study}, that our model achieves better accuracy with LFCC front--end as input for DeepFake data subsets. Most of DF generators are designed to be as realistic as possible from the perspective of human perception, which means that the methods often pay little attention to other artifacts that are not visible to humans. For this reason, it is relatively easy to detect classical DeepFake methods using appropriate signal features (e.g. LFCC).

In turn, the ASVspoof sub--dataset results are similar for both front--ends. We conclude that the mel--scale conversion reduces high--range frequencies and brings out the features of human hearing range. A human speech sample containing some artifacts in a high--frequencies band (out of hearing range) could be easily perceived by the automatic speaker recognition system, which usually works on a wider range of frequencies. Due to this reason, advanced spoofing datasets do not contain high--range frequency artifacts.
We could see that the lowest EER results out of sub--datasets are received for WaveFake. We conclude that the datasets are generated using similar methods. There are 7 architectures, while 6 of them are GANs. Moreover, some methods are interdependent (e.g. MelGAN, MelGAN Large), thus only 4 of them are unique. For FakeAVCeleb dataset, there is only one method used, thus the EER results are very similar for all folds. This indicates that other attacks do not have any impact on the SV2TTS method detection. 

We observe the highest variation of results for the ASVspoof sub--dataset. It contains the highest number of different attacks, split into train, test and eval subsets in a different way for every fold. Two folds with relatively worse results contain very selective validation subsets, including attacks that are distant from the train subsets. These results are significantly better for the fold 3 --- modifications from the train subset cover attacks from the test and eval subsets. Following Figure~\ref{fig:stability_train_test} we could observe that DeepFake detection training stability is generally low. With a constant and stable increment in accuracy for the train subsets, we could see large fluctuations in accuracy for the test subsets, which indicates that training of a DF detection model is not generally stable due to the use of multiple methods/attacks. However, the stability is significantly better for LFCC front--ends.

\section{Conclusion}

In this paper, we address topics of generalization and stabilization across Audio DeepFake detection models. We proposed a novel database --- Attack Agnostic Dataset --- a combination of 3 DeepFake and spoofing databases. We introduced the attack agnostic approach to training which, by disjoint division of the attacks, allows to address the aforementioned issues. Our work extensively analyzes and compares generalization and stabilization among currently used detection models, including the influence of different front--ends. We performed an ablation study, concluding that due to scaling of the frequencies in mel--based front--ends, linear front--ends are more suitable for detection.
We proposed a solution based on the LCNN model, which using LFCC front--end and mel--spectrogram provided better generalization and stabilization and overall performance in relation to commonly used LFCC. As a result, we achieved a significantly lower standard deviation that does not exceed 0.64\% on all folds. Furthermore, we decreased the EER of the baseline architecture by nearly 5\% by enhancing the input with a spectrogram.

\paragraph*{\uppercase{Acknowledgements}}
This work is partially supported by Polish National
Science Centre --- project UMO-2018/29/B/ST6/02969.

\bibliographystyle{IEEEtran}
\bibliography{mybib}

\end{document}